\documentclass[a4paper,11pt]{article}
\usepackage{pos}

\title{High-Energy Neutrino Production in Clusters of
	Galaxies
}
 \ShortTitle{Neutrinos production in Clusters}

\author*[a]{Saqib Hussain}
\author[b]{Rafael Alves Batista}
\author[c]{Elisabete de Gouveia Dal Pino}
\author[d,e]{Klaus Dolag}

\affiliation[a,c]{Institute of Astronomy, Geophysics and Atmospheric Sciences (IAG), University of S\~ao Paulo (USP), S\~ao Paulo, Brazil}

\affiliation[b]{Radboud University Nijmegen, Department of Astrophysics/IMAPP, 6500 GL Nijmegen, The Netherlands}
\affiliation[d]{University Observatory Munich, Scheinerstr. 1, 81679 Munchen, Germany}
\affiliation[e]{Max Planck Institute for Astrophysics, Karl-Schwarzschild-Str 1, 85741 Garching, Germany}

\emailAdd{s.hussain@usp.br}
\emailAdd{r.batista@astro.ru.nl}
\emailAdd{dalpino@iag.usp.br}

\abstract{In this work we compute the contribution from clusters of galaxies to the diffuse
	neutrino background. Clusters of galaxies can potentially produce cosmic rays
	(CRs) up to very-high energies via large-scale shocks and turbulent acceleration.
	Due to their unique magnetic-field configuration, CRs with energy $\leq 10^{17}$ eV can be trapped within these structures over cosmological time scales, and generate
	secondary particles, including neutrinos and gamma rays, through interactions
	with the background gas and photons. We employ three-dimensional cosmological
	magnetohydrodynamical simulations of structure formation to model the turbulent
	intergalactic medium. We use the distribution of clusters within this cosmological
	volume to extract the properties of this population. We propagate CRs in this
	environment using multi-dimensional Monte Carlo simulations across different
	redshifts (from $z \sim 5 \; \text{to} \; z =0$), considering all relevant photohadronic, photonuclear, and hadronuclear interactions. We also include the cosmological evolution of the CR sources. 
	We find that, for CRs injected with a spectral index
	$1.5 - 2.7$ and cutoff energy $E_{max} = 10^{16} -  10^{17}$ eV, clusters contribute to a substantial
	fraction to the diffuse flux observed by the IceCube Neutrino Observatory, and
	most of the contribution comes from clusters with $M > 10^{14} \;  M_{\odot}$ and
	redshift $z < 0.3$.}

\FullConference{37$^{\rm{th}}$ International Cosmic Ray Conference (ICRC 2021)\\
		July 12th -- 23rd, 2021\\
		Online -- Berlin, Germany}

\begin{document}
\maketitle

\section{Introduction}

The origin of high-energy neutrinos is not yet known, but their isotropic distribution as reported by the IceCube observations suggests they are predominantly extragalactic~\cite{Aartsen}. They can be produced by different classes of sources at different scales such as starburst galaxies, active galactic nuclei (AGNs)~\cite{fang2018, Anchordoqui, khaili2016},  supernova remnants \cite{senno2015}, and clusters of galaxies~\cite{hussain2019, murase2013}. The mechanisms whereby high-energy 
neutrinos are produced constitute one of the major mysteries in astrophysics today. These secondary particles can unveil the sources of ultra-high-energy cosmic rays (UHECRs) that produce them,
as they are not deflected by  magnetic fields while travelling through the intergalactic medium (IGM).  
The energy fluxes of UHECRs, high-energy neutrinos and gamma rays are comparable \cite{alves2019a, Ackermann2019}, which suggests that they can be explained by a single class of sources~\cite{fang2018}, like 
clusters or starburst galaxies~\cite{alves2019a}.
Clusters are 
attractive candidates for producing the UHECRs due to their large size ($\sim \text{Mpc}$) and magnetic-field strengths ($\sim \mu \text{G}$) \cite{kim2019}.
Clusters are likely formed by violent processes, like accretion and merging of smaller structures to larger ones~\cite{voit2005}. These processes release large amounts of energy ($10^{60}\; - \; 10^{65} \; \text{erg}$), part of which may accelerate CRs to very high energies via shocks and turbulence~\cite{Brunetti2020}.
The origin of the CRs with energies $E \lesssim 10^{17} \; \text{eV}$ is galactic~\cite{amato2018}, and those with $E \gtrsim 8 \times 10^{18} \; \text{eV}$ probably come from extragalactic sources~\cite{Aab2018}. However, the exact transition between the galactic and extragalactic CRs is not yet known\cite[see,][]{thoudam2016, kachelriess2019}.
Clusters can trap CRs of $E \lesssim 10^{17} \; \text{eV}$ for a time comparable to the age of universe~\cite{hussain2019}. During their confinement, CRs can interact with the gas present in the intra-cluster
medium (ICM) and produce secondary particles including gamma rays and neutrinos~\cite{kotera2009}. Hence, observations of these secondary particles may be used to constrain the properties
of CRs in clusters \cite{yoast2013,zandanel2015}.

There are many analytical and semi-analytical approaches being used to calculate the fluxes of secondary particle from clusters \cite[e.g.][]{kotera2009, murase2013}. In most of them the
model of the ICM is overly simplified. In~\cite{fang2016}, the authors estimated the flux of neutrinos from clusters by considering the isothermal gas distribution of the ICM and a Kolmogorov turbulent magnetic field, obtaining fluxes comparable to the IceCube measurements. In this work, we employ a more rigorous numerical approach to describe the propagation of CRs in the ICM.
Magnetic-field and density distributions are directly obtained from magnetohydrodynamics (MHD) large scale cosmological simulations of clusters of galaxies. Most important, we account for  the mass-dependent cluster properties since massive clusters ($M \geq 10^{15} \; M_{\odot}$) are less common than the lower mass ones ($M \leq 10^{14} \; M_{\odot}$). 
We also take into account the location of cosmic-ray sources inside the clusters.

To estimate the contribution of clusters to the flux of high-energy neutrinos, we follow the propagation of CRs and their
secondary particles in our MHD background simulation~\cite{dolag2005}
using the Monte Carlo code CRPropa~3\cite{Alves2016}. We account for all the relevant photohadronic, photonuclear, and hadronuclear interactions during the propagation, and compute the fluxes of CRs and neutrinos. In section~\ref{sec:MHD-cluster}, we discuss the MHD simulations of structure formation and properties of clusters. In section~\ref{sec:CRs-sim}, we describe the simulation setup for the propagation of CRs in ICM and relevant particle interactions. The resulting fluxes of CRs and neutrinos are discussed in section~\ref{sec:flux-CRs-Neu}, followed by the conclusions in section~\ref{sec:conclusions}.

\section{Background MHD Simulation and Cluster Properties}\label{sec:MHD-cluster}

We considered the cosmological 3D-MHD simulations from~\cite{dolag2005}. From these simulations we obtained the mass, temperature, density, and magnetic-field distributions of galaxy clusters, filaments, and voids. We considered seven snapshots with volume of sphere of radius $110 \; h^{-1} \; \text{Mpc}^3$ from the simulations at  redshifts 
$z = 0.01; \; 0.05; \; 0.2; \;0.5; \;0.9;\; 1.5;\; 5.0$. Each snapshot 
was divided into eight regions.
In these simulations, AGN feedback and star formation were not considered. The  cosmological parameters assumed are $h = H_0/(100 \; \text{km} \; \text{s}^{-1} \; \text{Mpc}^{-1})=0.7, \; \Omega_m = 0.3, \; \Omega_\Lambda=0.7$, and the baryonic fraction $\Omega_b/\Omega_m = 14\%$ (for more details see~\cite{hussain2021,dolag2005}). 

\begin{figure}
	\includegraphics[width=8cm]{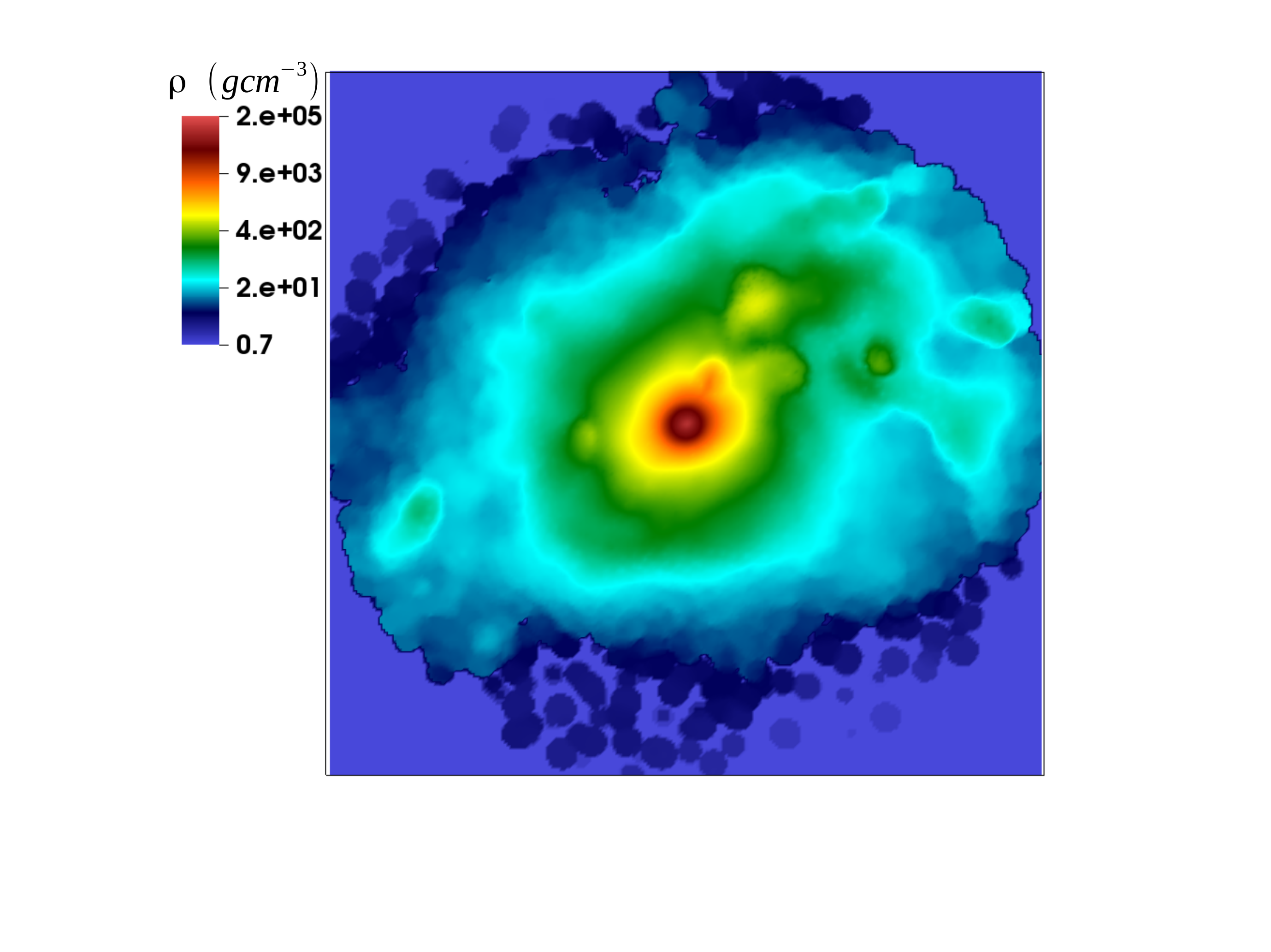} 
	\includegraphics[width=8cm]{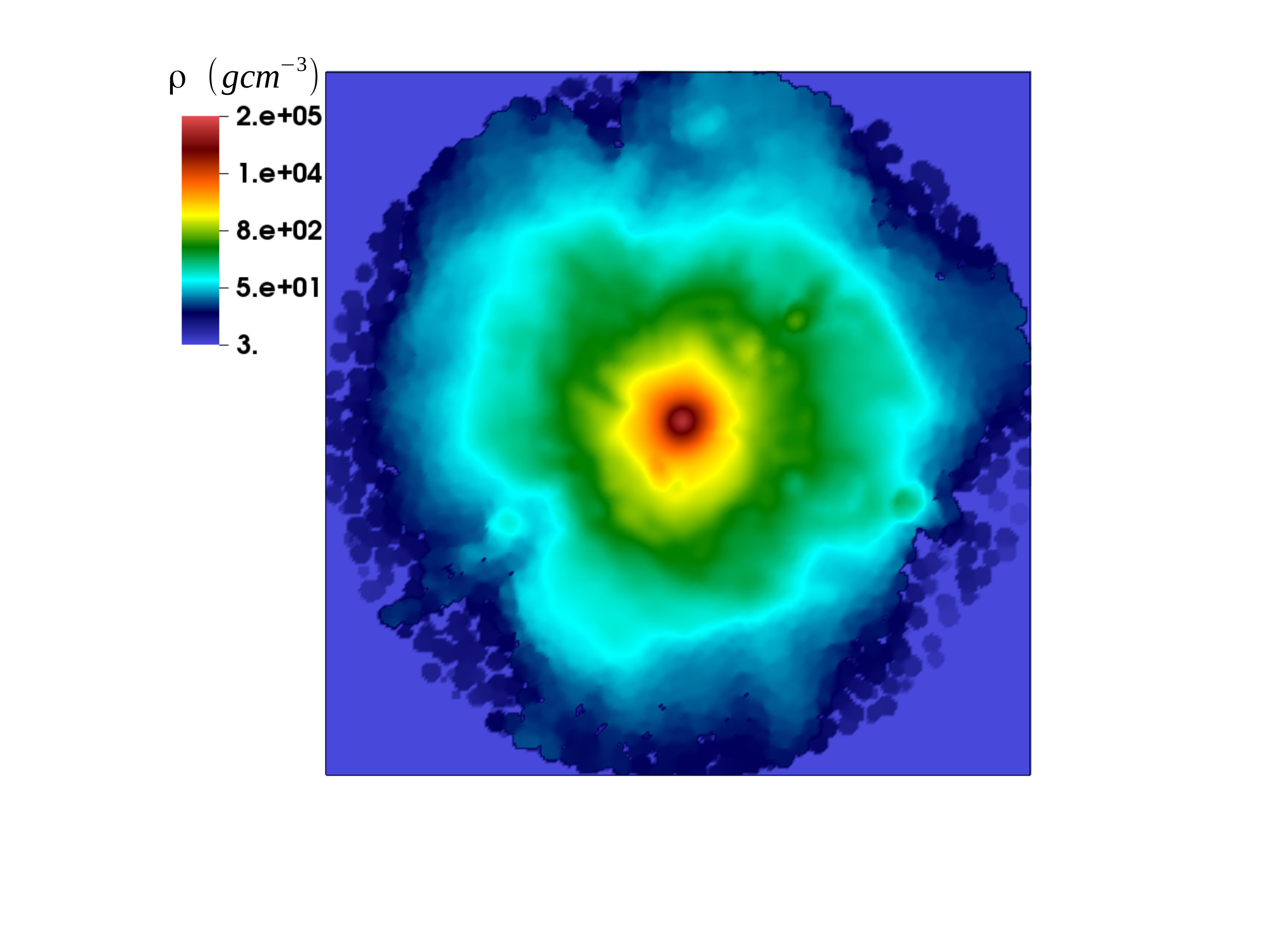}
	\caption{Maps showing the gas density for two clusters of masses $\simeq 10^{14} M_{\odot}$ (left panel) and  $\simeq 10^{15} M_{\odot}$ (right panel), at redshift $z=0.01$. (Extracted from ~\cite{hussain2021}).}
	\label{fig:Cluster-contour}
\end{figure}

\begin{figure}
	\includegraphics[width=6cm]{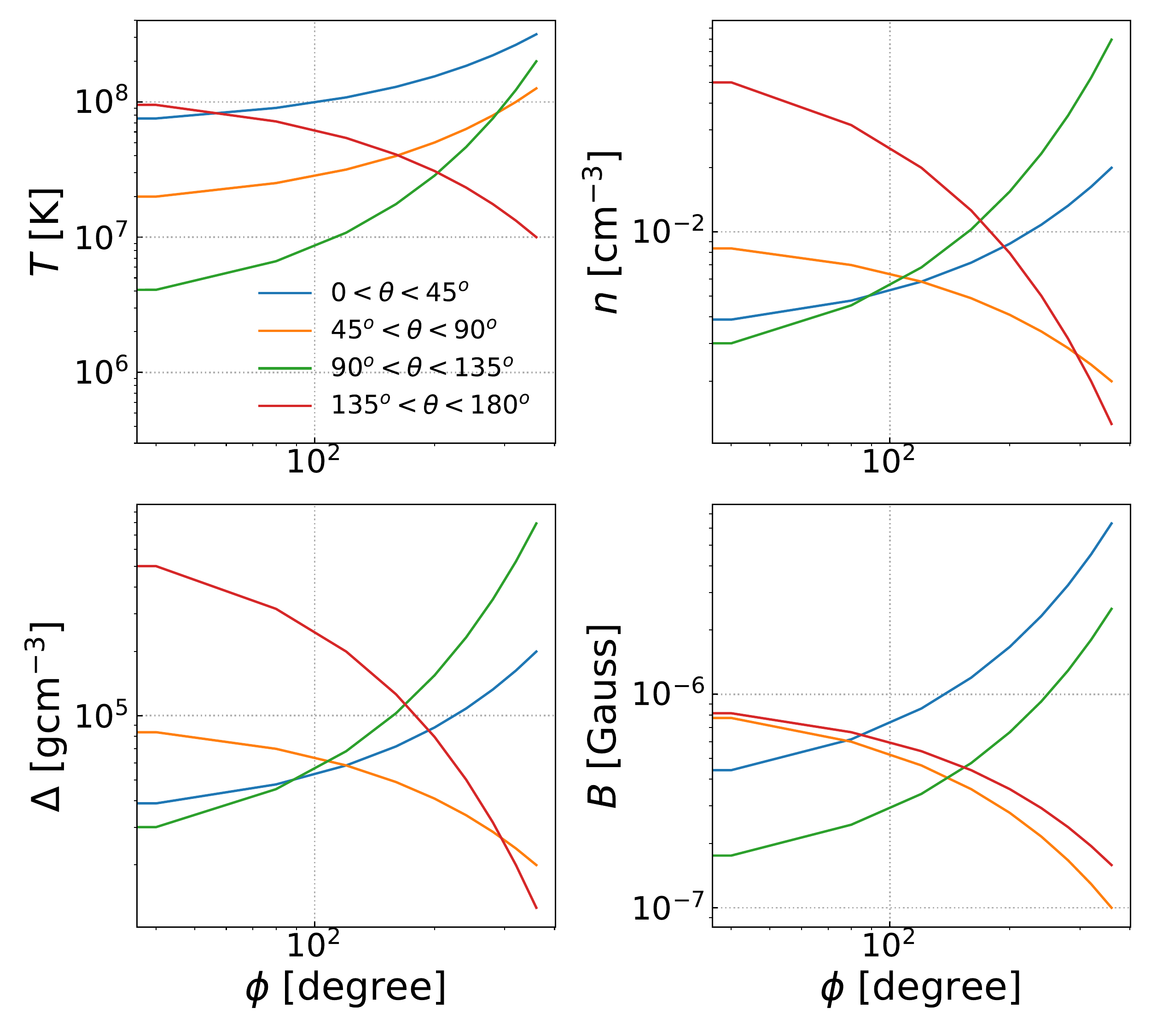}
	\includegraphics[width=8cm]{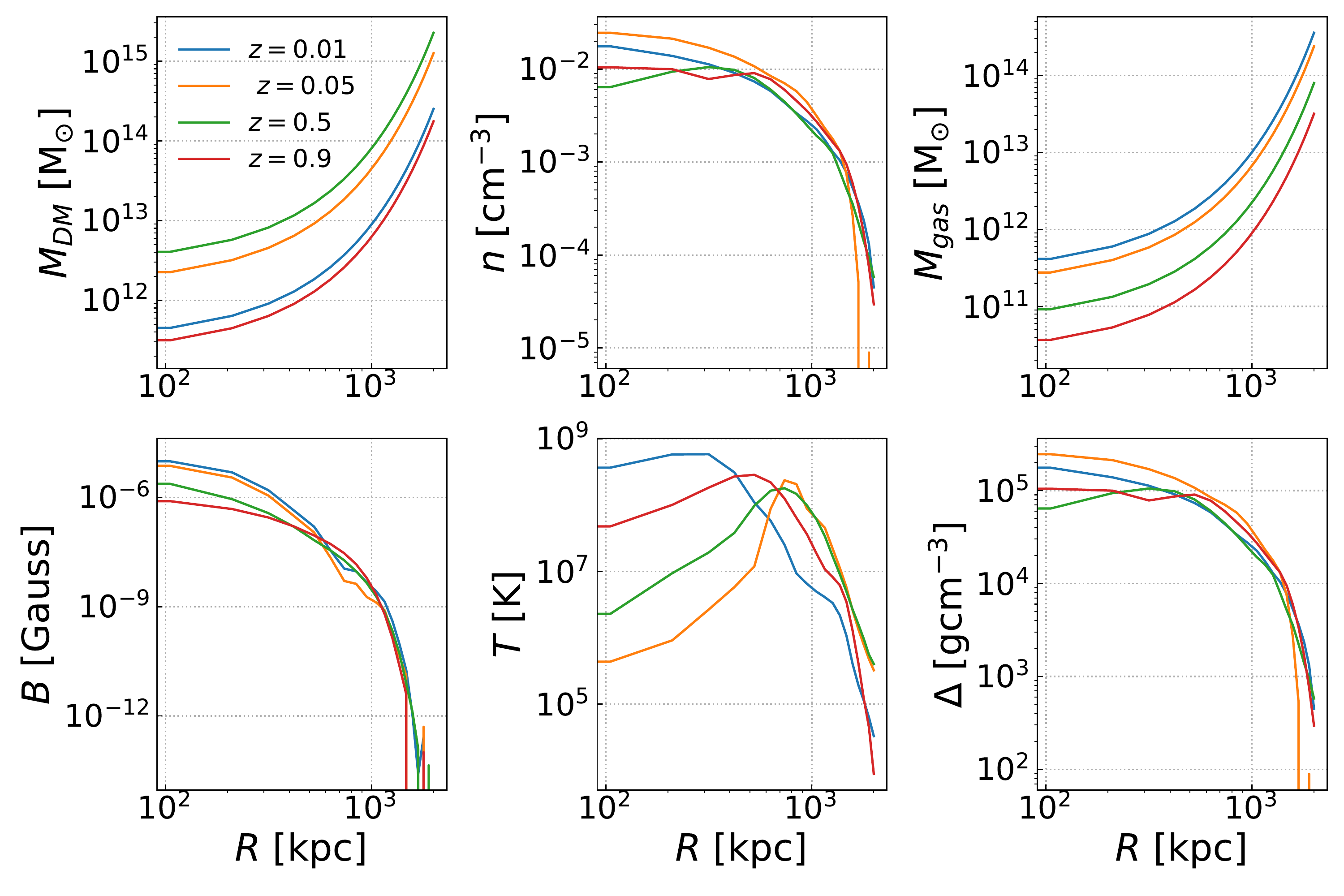}
	\caption{Left set of panels show the volume-averaged profiles of temperature, and number density (top), and overdensity and magnetic field (bottom) as a function of the azimuthal angle ($\phi$), for different zenith angles ($\theta$), within a radial distance $R= 300 \text{kpc}$ from the center. The right set of panels shows the dark-matter, gas number density, and gas mass (top); and magnetic field, temperature, and  overdensity (bottom), as a function of the radial distance from the center, for four redshifts indicated. Here the cluster has a mass of $M \simeq 10^{15} \; M_{\odot}$.
	(Extracted from ~\cite{hussain2021}).
	} 
	\label{fig:ClusterProp}
\end{figure}

For each redshift, we  identified  the clusters in the densest regions of the entire volume of the simulation and then selected five clusters with different masses in the range $10^{12} \leq M/M_{\odot} < 10^{16}$. 
These clusters were considered as  representative of all the clusters in the corresponding redshift. We then studied the propagation of CRs and computed the associated neutrino flux stemming from CR interactions with the ICM. 
In Fig.~\ref{fig:Cluster-contour}, we show the density maps for two clusters of masses $10^{14} \; M_{\odot}$ and $10^{15} \; M_{\odot}$, and  in in Fig.~\ref{fig:ClusterProp} the volume averaged profiles of the clusters with respect to dark-matter mass; gas number density; gas mass; magnetic field; temperature and  overdensity. Distributions of temperature, magnetic field, and  density are not spherically symmetric inside the clusters, as shown in Fig.~\ref{fig:Cluster-contour} and the left panel of Fig.~\ref{fig:ClusterProp}. To estimate the cluster masses, we integrated the baryonic and dark-matter densities within a volume of $(\sim 2 \text{Mpc})^3$, assuming  a spherical volume. We found that, while the mass is not substantially affected by the deviations from spherical symmetry, the emission pattern of the CRs and
corresponding secondaries change considerably. For more details about the magnetic-field profiles and the mass distribution of clusters, see~\cite{hussain2021}.

\section{Simulation Setup for Cosmic Rays}\label{sec:CRs-sim}

We employ the Monte Carlo code CRPropa~3~\cite{Alves2016} with
stochastic differential equations~\cite{merten2017} to study the diffusion of CRs in the ICM. The MHD simulations provide the background magnetic field, gas density, and temperature distributions of the ICM. We considered all relevant interactions during their propagation including photohadronic, photonuclear, and hadronuclear processes, namely photopion production, photodisintegration, nuclear decay, and adiabatic losses.
In addition to the CMB and EBL photon fields, there are interactions
of CRs with the gas of the ICM (proton-proton; $pp$); in fact, this is the dominant channel for neutrinos and gamma-ray production.
The bremsstrahlung radiation~\cite{rybicki2008} produced by the hot gas  ($T \sim 10^{6} - 10^{8} \; \text{k}$) of the ICM medium can also interact with the CRs.

In CRPropa~3, the CR transport is done using the Parker transport equation (see \cite{merten2017}). The regime of propagation of CRs, whether diffusive, semi-diffusive, or ballistic, depends upon their Larmor radius ($r_\text{L}= 1.08 E_{15}/B_{\mu\text{G}} \; \text{pc}$). In the case of clusters we are in the diffusive regime as the Larmor radius for CRs of energy $10^{14} \leq E/\text{eV} \leq 10^{19}$ is less than the 
typical cluster size ($ R_\text{cluster} \sim 1 \; \text{Mpc} $).
For instance, $r_\text{L} \sim 0.1 \; \text{kpc}$ for CRs with energy $\sim 10^{17} \; \text{eV}$ in a cluster
of mass $\sim 10^{14} \; M_{\odot}$ with a magnetic field $\sim \mu\text{G}$, which is evidently much less than the size of the cluster ($\simeq 2 \; \text{Mpc}$). 
The trajectory length of the CRs inside the clusters is $\sim 10^3 \; \text{Mpc}$, such that their confinement time  is $t_\text{con}\sim 10^3 \; \text{Mpc}/c \sim t_\text{H}$ (being the latter, the Hubble time). Therefore, CRs with $E \lesssim 10^{17} \; \text{eV}$ can be trapped inside clusters for cosmological times. Note, however, that the spectrum of CRs escaping from a cluster is highly dependent on the cluster mass and its magnetic-field profile. 



\begin{figure}
    \centering
	\includegraphics[width=0.49\textwidth]{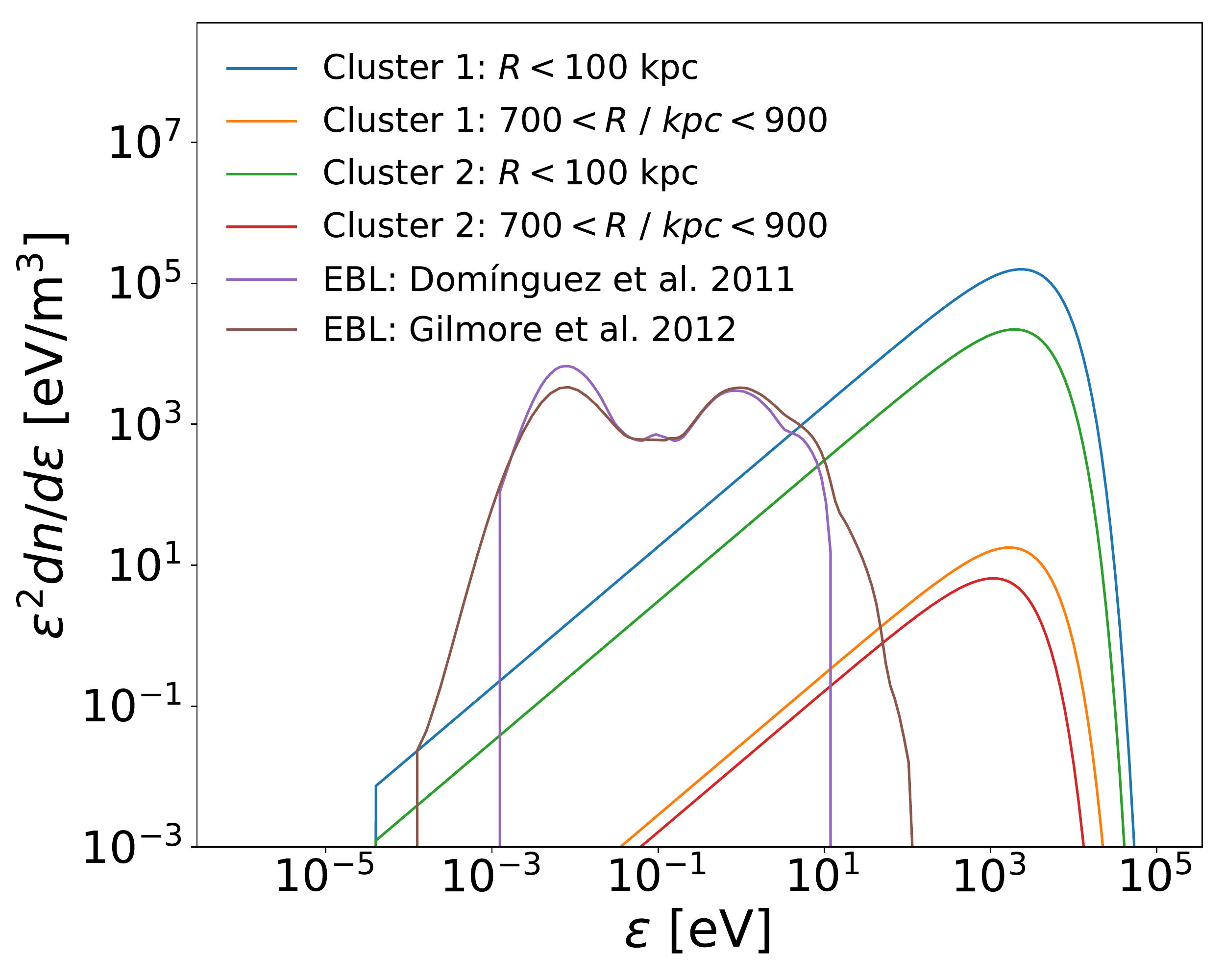}
	\includegraphics[width=0.49\textwidth]{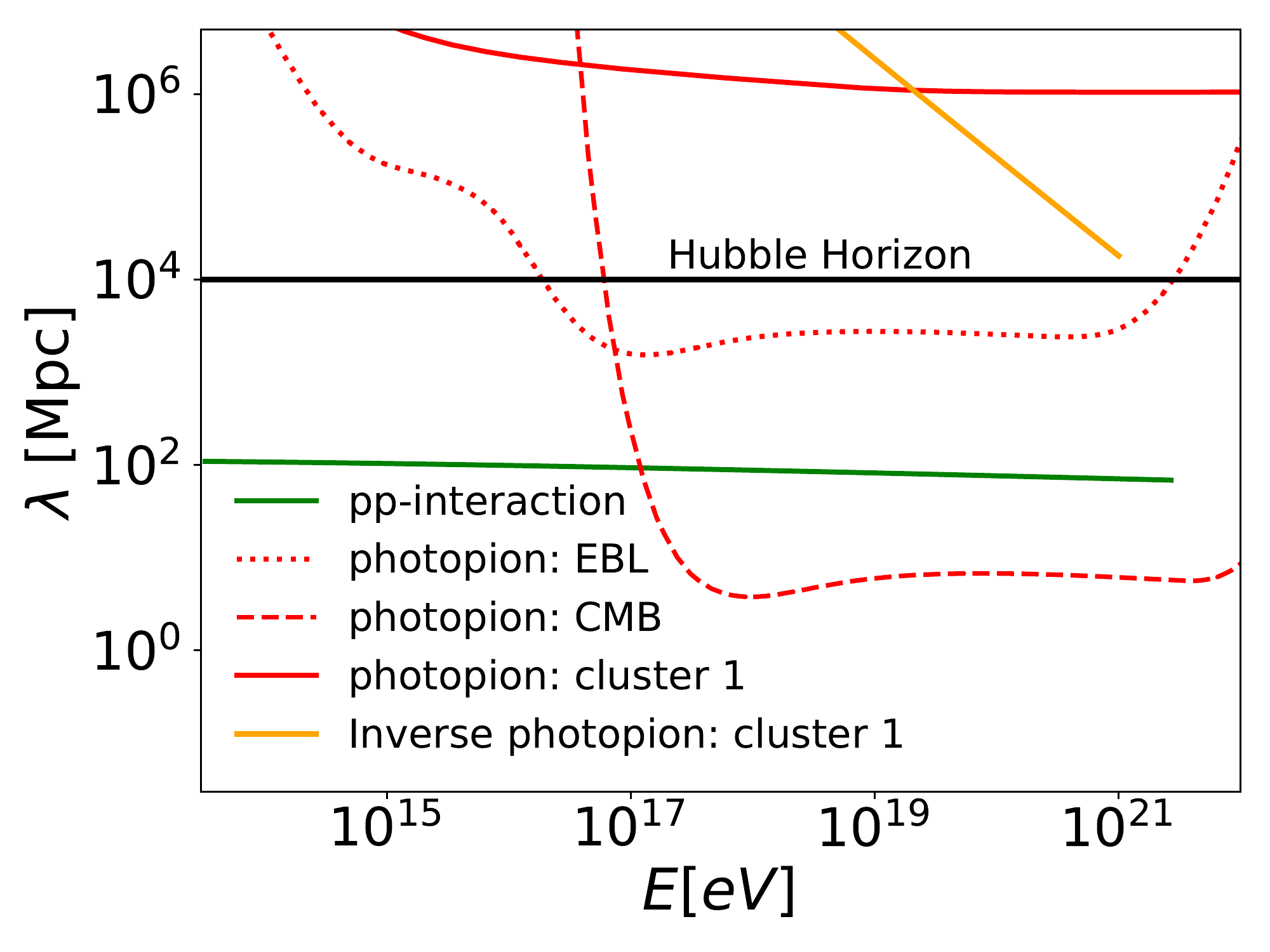}
	\caption{The left panel shows the energy density of different photon fields, namely the CMB, the EBL, and the average bremsstrahlung radiation of the ICM in  representative regions of two clusters. Cluster$\; 1$ has mass $10^{15} \; M_{\odot}$, and Cluster$\;2$ , $10^{14} \; M_{\odot}$.
	The right panel shows the mean free path ($\lambda$) for the CR interactions that produce neutrinos, including photopion production in the bremsstrahung photon field (red solid line), in the CMB (red dashed line), and in the EBL (red dotted line), as well as hadronuclear interactions (green) calculated within a sphere of radius $r = 100$ kpc around the center of the massive cluster ($\sim 10^{15} \; M_{\odot}$ shown in Fig.~\ref{fig:Cluster-contour}). The sub-dominant interaction of high-energy gamma rays with the local gas of the ICM is also shown for comparison (orange line). The thick black line represents the Hubble horizon. (From ~\cite{hussain2021}).
	}
	\label{fig:EBL-MFP-Traj}
\end{figure}

We calculated the bremsstrahlung photon field of the ICM and compared it with EBL in the left panel of Fig.~\ref{fig:EBL-MFP-Traj} (for details see \cite{hussain2021}). The bremsstrahlung radiation is dominant at X-rays energy, but only near the center of the clusters, while the EBL dominates at infrared and optical wavelengths mainly.

To see which CR interactions are important to produce neutrinos inside clusters, we calculated the mean free path (MFP) considering the gas of the ICM ($pp$-interactions), the bremsstrahlung radiation field, the EBL and CMB, as shown in the right panel of Fig.~\ref{fig:EBL-MFP-Traj}. 
The MFPs for $p\gamma$ interactions with the CMB and for $pp$ interactions with the gas are much smaller than those for interactions with the EBL and the bremsstrahlung field, such that the former is more likely to occur. Moreover, interactions of high-energy gamma rays with the local gas of the ICM  (inverse photopion production) and CR interactions with bremsstrahlung radiation (cluster $1$) can be neglected because their MFP \textbf{are} larger than the Hubble radius. 
 

\section{Flux of Cosmic-rays and Neutrinos}\label{sec:flux-CRs-Neu}

Cosmic-ray protons were injected isotropically inside the clusters assuming 
a power-law spectrum $E^{-\alpha}$ with an exponential cut-off energy $E_{\text{max}}$. We considered different values for the maximum energy, $E_{\text{max}} = 5\times10^{15} - 10^{18}$~eV, and for the spectral index $\alpha \simeq 1.5 - 2.7$. In our background simulations, the AGN feedback and star formation were not considered. To account for their contribution in the spectrum, we adopted analytical expressions for the evolution of these objects, since they can produce the high-energy CRs of interest, following~\cite{alves2019b, hussain2019}.
The luminosity of CRs from our background simulations is several orders of magnitude smaller than the luminosity of observed clusters. To convert the code units of the simulations to physical units we compared the cluster luminosity with the simulated CRs and obtained the normalization factor (see \cite{hussain2021} for details).

\begin{figure}
    \centering
	\includegraphics[width=0.49\textwidth]{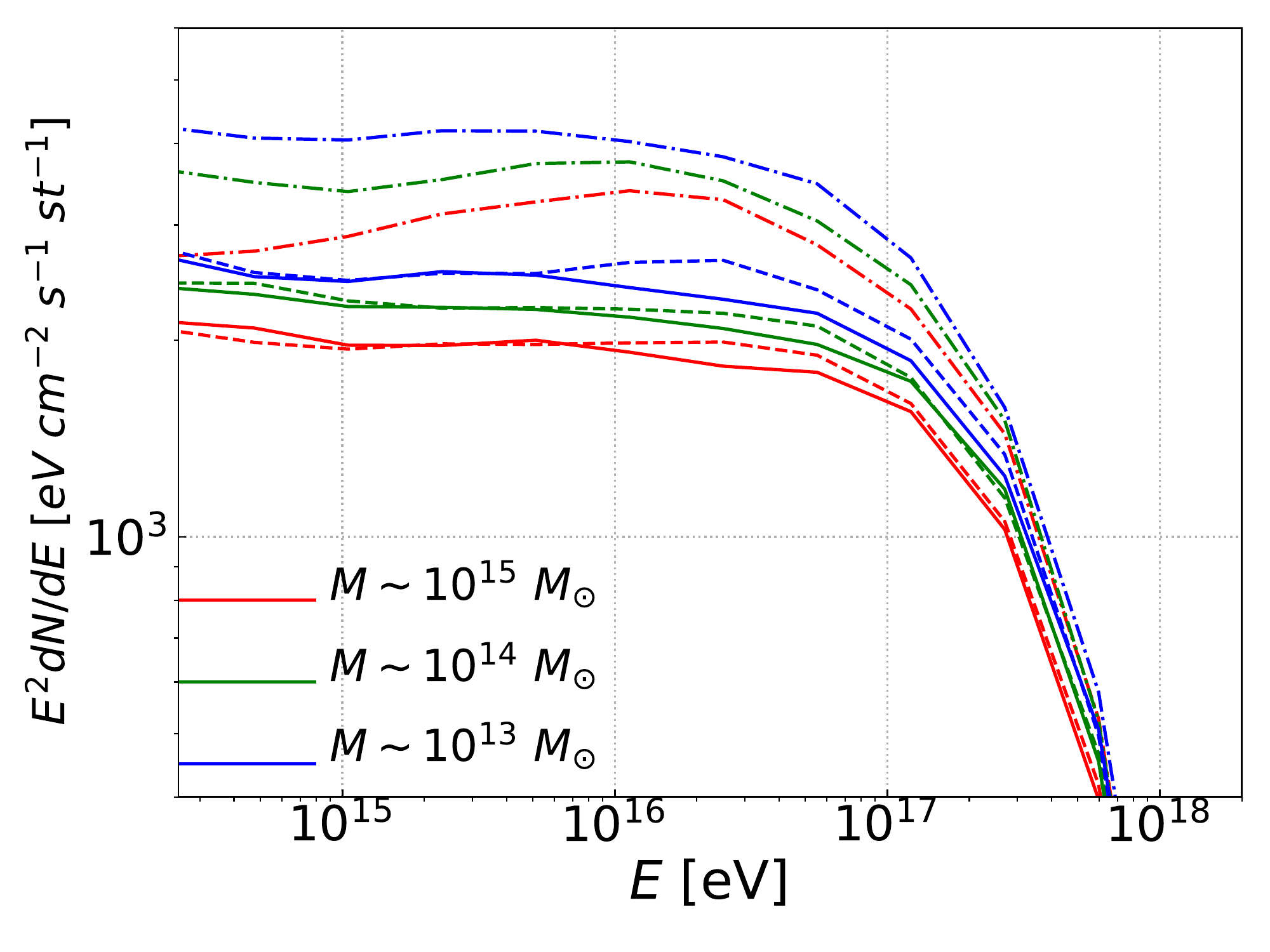}
	\includegraphics[width=0.49\textwidth]{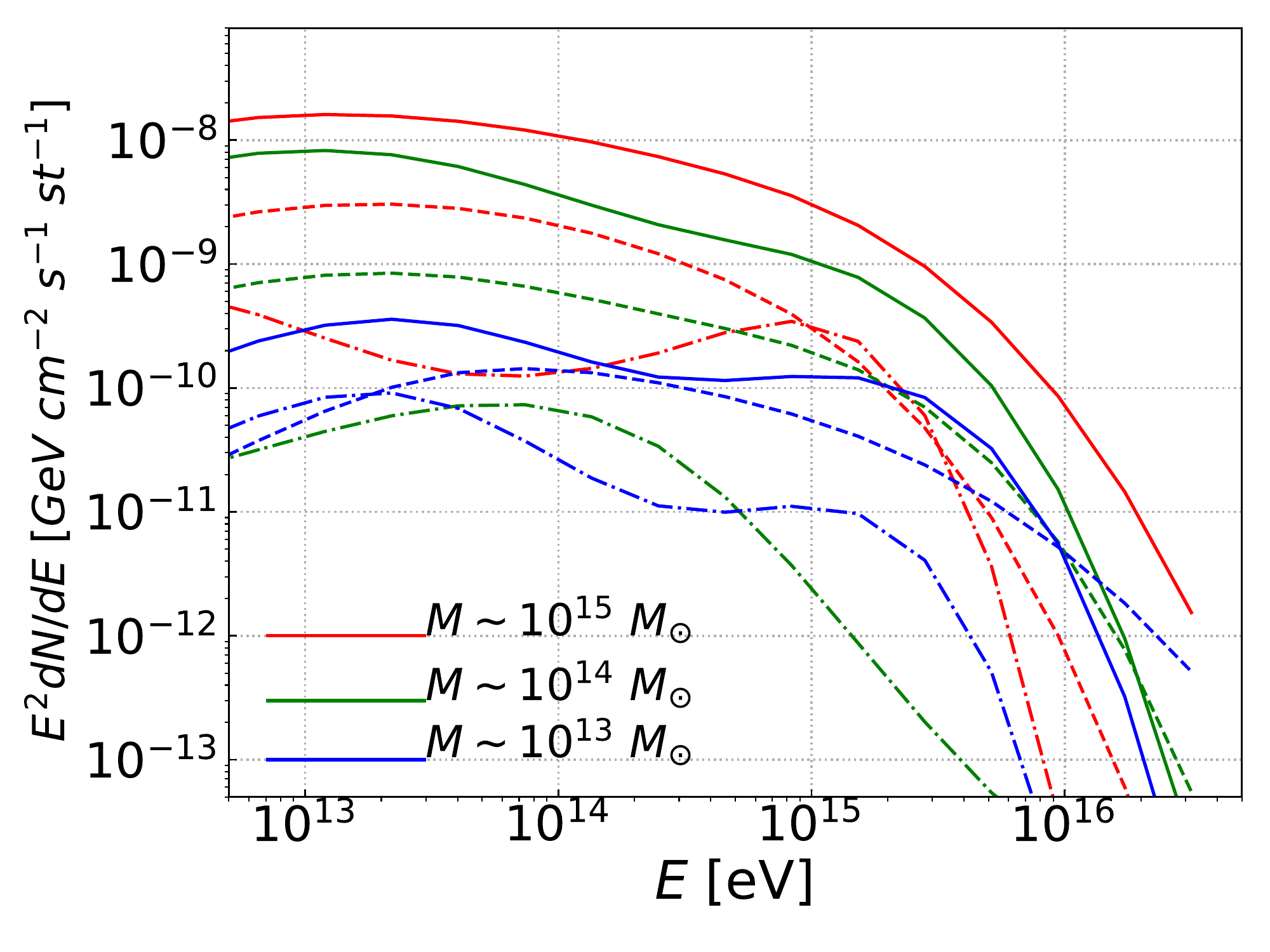}
	\caption{
	Fluxes of CRs (left panel) and neutrinos (right panel) for individual clusters of different masses: $M\sim 10^{15}$ (red);  $10^{14}$ (green); and $M\sim 10^{13}~M_{\odot}$ (blue lines). Sources located at different positions in the cluster are represented by different line styles: center of the cluster (solid), $300$~kpc  (dashed)  and $1$ Mpc away from the center (dash-dotted lines). The flux is computed at the edge of the clusters. The spectral parameters are $\alpha=2$ and $E_{\text{max}} = 5\times 10^{17}$~eV. It is assumed that $2\%$ of the luminosity of the clusters is converted into CRs.
	(From ~\cite{hussain2021}).
}
	\label{fig:CR-neu_Offset}
\end{figure}

In order to study the spectral dependence with position, we selected different injection points for CRs inside clusters of different mass. The resulting fluxes are shown in the left panel of Fig.~\ref{fig:CR-neu_Offset}.  
One can see that the CRs injected at the outskirts of the clusters (at $1 \; \text{Mpc}$) can escape easily compared to other injection points closer to the  cluster center. This follows from the fact that the magnetic field and gas density there are low in comparison with the central regions~\cite{dolag2005}.
Furthermore, we found a significant suppression in the flux of CRs at $E \gtrsim 10^{17}\; \text{eV}$ thus indicating the trapping of low-energy CRs inside clusters \cite{hussain2019, hussain2021}.
The spectra of CRs have been collected  in a sphere of radius of 2 $\text{Mpc}$ centred at the cluster. The all-flavour  neutrino fluxes are also computed for these same observers.

\begin{figure}
	\includegraphics[width=5cm]{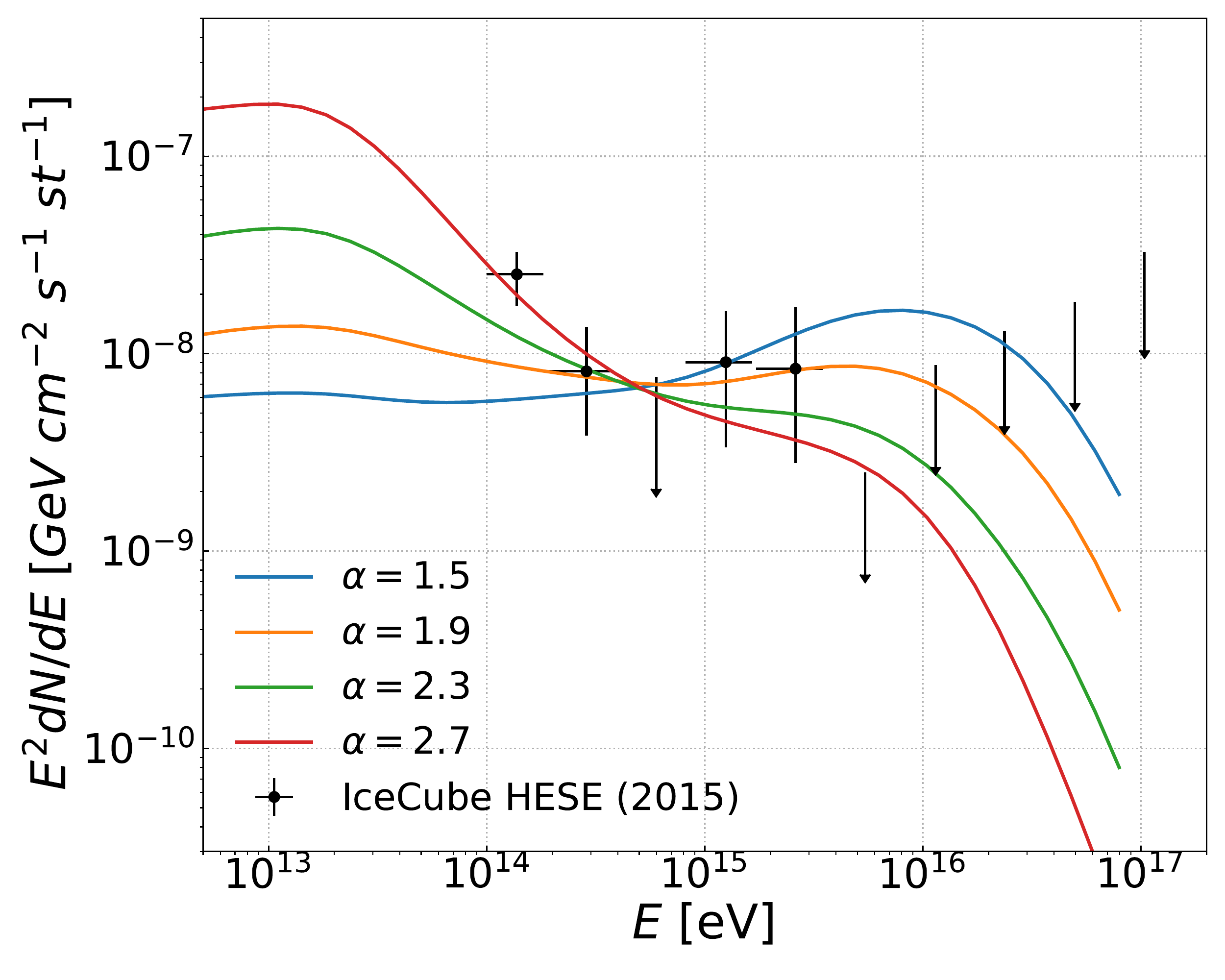}
	\includegraphics[width=5cm]{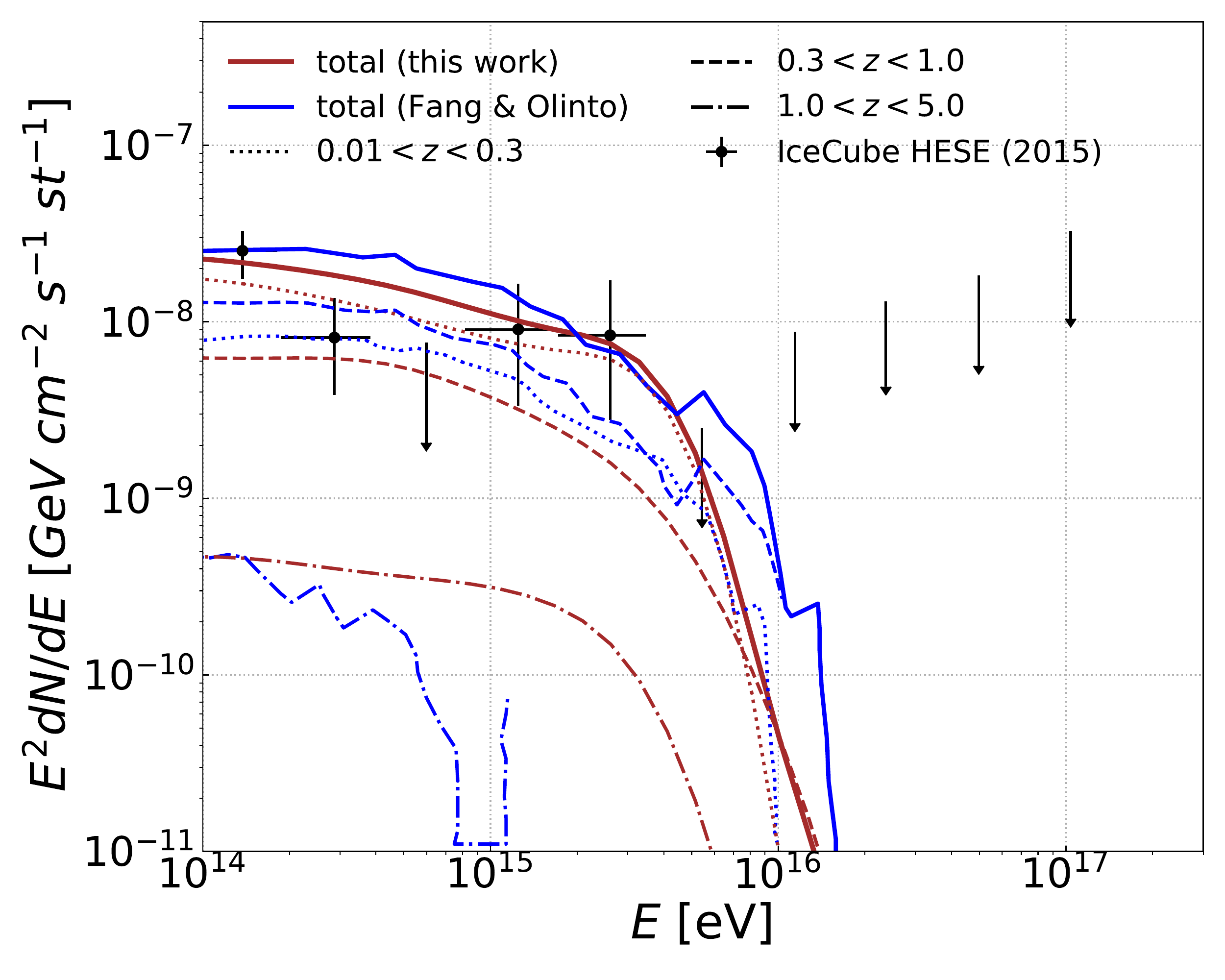}
	\includegraphics[width=5cm]{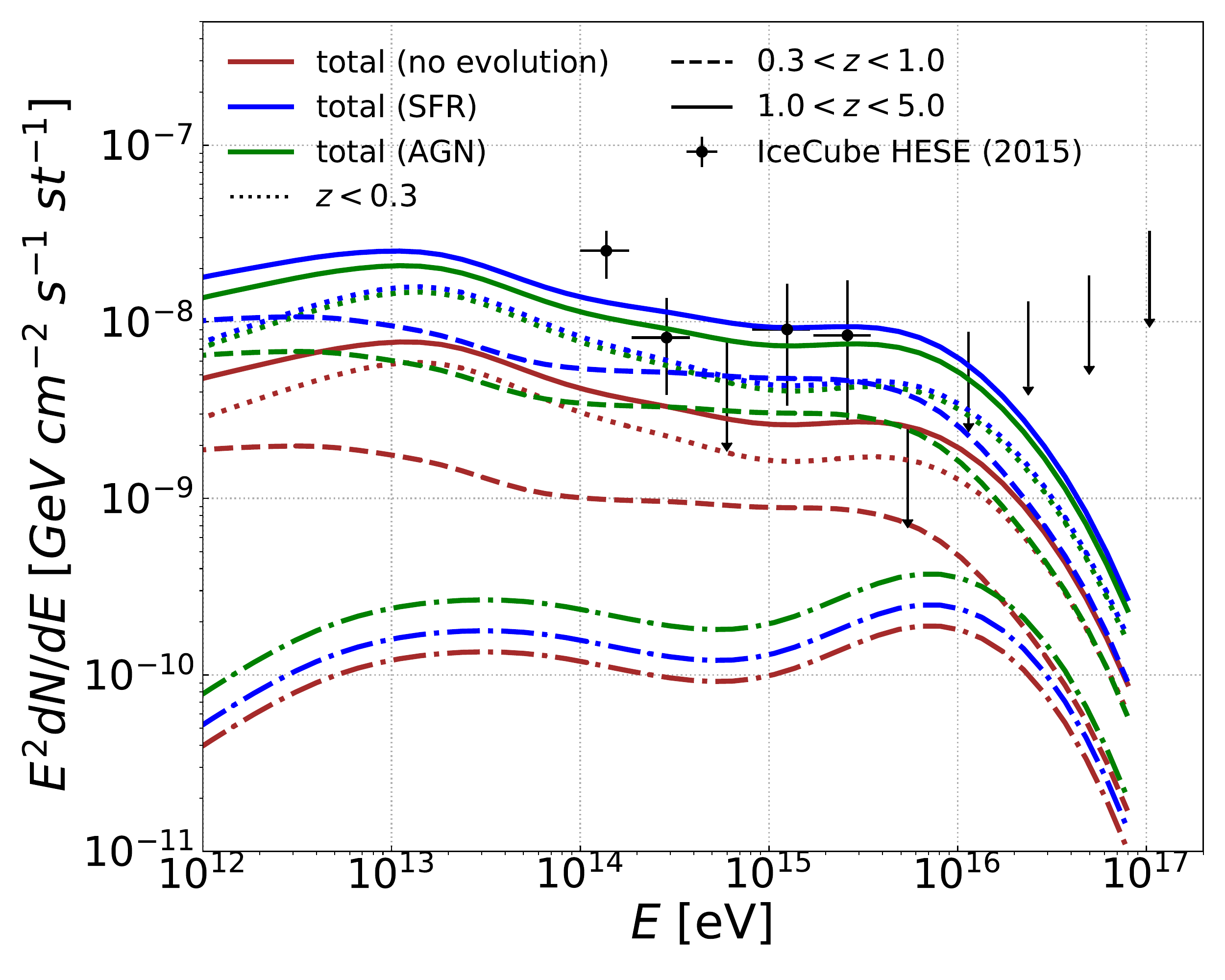}
	\caption{
	The left panel shows the total spectrum of neutrinos for different injected CR spectra with $\alpha = 1.5$, $1.9$, $2.3$, $2.7$ and  $E_\text{max} = 5\times 10^{17}~\text{eV}$. The middle panel shows the calculated neutrino spectra for different redshift ranges: $z < 0.3$ (dotted lines), $0.3 < z < 1.0$ (dashed), and $1.0 < z < 5.0$ (dash-dotted lines). The solid blue and brown lines correspond to the total spectrum in \cite{fang2016}, and in this work, respectively. The CR injection in this figure is for $\alpha=1.5$ and $E_\text{max} = 5 \times 10^{16} \; \text{eV}$. The right panel shows the neutrino spectra considering different sources evolution: SFR (blue), AGN (green), and no evolution (brown lines). The CR injection spectrum has parameters $\alpha=2$ and $E_\text{max} = 5 \times 10^{17} \; \text{eV}$.
		(From ~\cite{hussain2021}).}
	\label{fig:neu-all}
\end{figure}

The production of neutrinos inside the clusters occurs mainly due to photopion production and $pp$-interaction, as one can infer from the MFPs shown in Fig.~\ref{fig:EBL-MFP-Traj}. We observed less neutrino production for the injection of CRs at $1 \; \text{Mpc}$ away from the cluster center. Also, we found that more massive clusters produce more neutrinos, as shown in the right panel of Fig.~\ref{fig:CR-neu_Offset}. In Fig.~\ref{fig:neu-all} we plotted the total flux of neutrinos from the whole population of clusters obtained from the background 3D-MHD simulation in the redshift range $0.01 \leq z \leq 5.0$. In order to match the observed IceCube measurements, we plotted the flux of neutrinos for different spectral indices ($\alpha$) and cut-off energies $E_\text{max} = 5\times 10^{17}\; \text{eV}$ in the left panel of Fig.~\ref{fig:neu-all}. All choices of $\alpha$ roughly match with the IceCube data. We assumed that $3\%$ of the kinetic energy of the clusters are converted into CRs~\cite{hussain2021, fang2016}. In the middle panel of Fig.~\ref{fig:neu-all}, we compared our results with Fang \& Olinto (2016)~\cite{fang2016}, obtaining a reasonable agreement. The main difference between our results and theirs is that in our case the major contribution to the total neutrino flux  comes from the redshift interval $0.01 < z < 0.3$, whereas in their case this would be from $0.3 < z < 1.0$. This difference is possibly due to the more simplified modeling of the background distribution of clusters in \cite{fang2016}, specially for the lower mass group ($10^{12} - 10^{14} M_{\odot}$ ) at high redshifts ($z > 1$).

We illustrate the impact of the redshift evolution of the CR sources on the flux of high-energy neutrinos in the right panel of Fig.~\ref{fig:neu-all}. We compared the results for different \cite{fang2016} sources evolution such as AGN and SFR, with that with no evolution. We find that the production of neutrinos is higher for the AGN and SFR cases, which is expected due to the redshift dependence of their emissivity. The effect of AGN evolution is dominant at high redshift ($1.5 \leq z \leq 5.0$) as compared to the evolution of SFR.

\section{Discussion and Conclusions}\label{sec:conclusions}

We calculated the contribution of clusters to the diffuse high-energy neutrino background using 3D-MHD cosmological simulations. From these simulations we obtained the relevant quantities that describe the ICM such as the magnetic field, the gas density, and the temperature, which depend on the cluster mass. 
CRs with energy $E\leq 10^{17}\; \text{eV}$ can be trapped inside the magnetic field of massive clusters ($\sim 10^{15}\; M_\odot$) and, as a consequence, they are more likely to interact with the
gas of the ICM and produce more high-energy neutrinos.
Conversely, the flux of neutrinos is lower for less massive clusters because the magnetic-field strength and the gas density are lower. 
Thus, we conclude  that the neutrinos above PeV energies are produced mainly in massive clusters because the CR interactions with the gas are more rare in clusters of mass $M < 10^{14} \; M_{\odot}$.

In order to try to  reproduce the IceCube date we considered different injection CRs spectral indices ($\alpha \simeq 1.5 - 2.7$) and cut-off energies  ($E_{\text{max}} = 5\times10^{15} - 10^{18}$ eV), as well as the CR sources evolution (AGN, SFR, no evolution) in Fig.~\ref{fig:neu-all}.
Our results suggest  that the clusters  can contribute to a sizable fraction of the diffuse neutrinos flux, if the CRs composition is mainly protons.
We also showed that the redshift evolution of the CR sources can increase the production of neutrinos inside clusters. To better assess how the redshift evolution of the CR sources affects the total neutrino flux from clusters,  more realistic cosmological simulations will be required, to account for AGN and star formation feedback from galaxies \cite{barai2019}.


\section*{Acknowledgements}
We are grateful for the support of the Brazilian agencies FAPESP (grant 2013/10559-5 \& 17/12828-4) and CNPq (grant 308643/2017-8). RAB acknowledges the financial support from the Radboud Excellence Initiative.

\end{document}